# Question on Conditional Entropy

WANG Yong

(School of Computer and Control, GuiLin University Of Electronic Technology ,Guilin City, Guangxi Province , China, 541004)

*Abstract*—**The problems of conditional entropy's definition and the formula to compute conditional entropy are analyzed from various perspectives, and the corrected computing formula is presented. Examples are given to prove the conclusion that conditional entropy never be increased is not absolute, thus the representation that information is to decrease uncertainty in the definition of information is not absolutely correct.**

## I. Introduction

Shannon's information theory has far-reaching effects on the development of telecommunication technology[1]. But his theory is not adaptable in many aspects of our daily life. A lot of scholars, even Shannon himself, realized the limitations of Shannon's information theory. Lots of problems were pointed out [2-4]. I have pointed out in other papers that the definition of information does not take the reliability of information into consideration and given a new definition to information [5]. This paper makes further attempts to explain why Shannon's information theory does not suit reality from certain angles. Shannon introduced the concept of entropy into information theory, which originally came from thermodynamics. In thermodynamics, entropy will never be decreased in any known physical evolution of the isolated system. However, in Shannon's information theory, conditional entropy is never increased. The author found that the conditional entropy may be increased when studying the perfect security of the one-time-pad [6]. This paper aims to seek out the source of the mistakes about conditional entropy through examples and analyses on the limitations of Shannon's proof.

## II. Literature review on conditional entropy

In the literature [1], Shannon provided the definition to entropy of the joint events and conditional entropy, produced the formula, and made some statements and reasoning as follows:

Suppose there are two events, x and y, in question with m possibilities for the first and n for the second.

Let p(i; j) be the probability of the joint occurrence of i for the first and j for the second. The entropy of the joint event is

$$H(x, y) = -\sum_{i,j} p(i, j) \log p(i, j)$$

$$H(x) = -\sum_{i,j} p(i,j) \log \sum_{j} p(i,j)$$

$$H(y) = -\sum_{i,j} p(i,j) \log \sum_{i} p(i,j)$$

It is easily shown that

$$H(x,y) \leq H(x) + H(y)$$

With equality only if the events are independent (i.e., p(i; j) = p(i)p( j)). The uncertainty of a joint event is less than or equal to the sum of the individual uncertainties.

For any particular value i that x can assume there is a conditional probability $p_i(j)$ that y has the value j. This is given by

$$p_i(j) = \frac{p(i,j)}{\sum_{j} p(i,j)}$$

We define the conditional entropy of y, $H_x(y)$ as the average of the entropy of y for each value of x, weighted according to the probability of getting that particular x. That is

$$H_x(y) = -\sum_{i,j} p(i,j) \log p_i(j) \qquad (1)$$

This quantity measures how uncertain we are of y on the average when we know x. Substituting the value of $p_i(j)$ we obtain

$$H_x(y) = -\sum_{i,j} p(i,j) \log p_i(i,j) + \sum_{i,j} p(i,j) \log \sum_{j} p(i,j)$$

$= H(x,y) - H(x)$

or

$H(x,y) = H(x) + H_x(y)$

The uncertainty (or entropy) of the joint event x;y is the uncertainty of x plus the uncertainty of y when x is known.

For

$$H(x,y) \leq H(x) + H(y)$$

$H(x) + H(y) \geq H(x,y) = H(x) + H_x(y)$

Hence

$H(y) \geq H_x(y)$.

Shannon made the conclusion that the uncertainty of y would be never increased by knowledge of x. and it would be decreased unless x and y were independent events, in which case it was not changed.

## III. Limitations Analyses on the proof of conditional entropy

According to the above description that the uncertainty of y is never increased by knowledge of x, Hx(y) should be equal to the entropy of y when x occurs.

To get the entropy of y when x occurs, we can compute the probability that y has the value j when x occurs. That is

$$p_x(j) = \sum_i (p_i(j) p(i))$$

**Then we have**

$$H_x(y) = -\sum_j p_x(j) \log p_x(j) = -\sum_j (\sum_i (p_i(j) p(i))) \log \sum_i (p_i(j) p(i)) \quad (2)$$

According to formula (2), it can be seen that the impact of x on the probability of y is random, which means the uncertainty of y can either be decreased or increased. But according to formula (1), **we can infer that** $H(y) \geq H_x(y)$.

The following table which shows the joint probability distributions of x and y is to illustrate the increase of conditional entropy.

Table 1 the joint probability distributions of x and y

| x \ y | y=0 | y=1 |
|---|---|---|
| x =0 | 0. 2 | 0. 3 |
| x =1 | 0. 1 | 0. 4 |

The prior probability of y is that p(y=0)=0.3 and p(y=1)=0.7.

Assumed that p(x=0)=1 and p(x=1)=0, Then the posterior probability of y is that p(y=0|x)=0.4 and p(y=1|x)=0.6. It can be inferred that the uncertainty of y is increased by knowledge of x.

The problem with formula (1) is that it computes the average of the entropy of y for each value of x, weighted according to the probability of getting that particular x. With this method, what is got is not the entropy of y when x occurs, but the weighted average of the entropies. Furthermore, that the value obtained from formula (1) is no more than H(y) cannot educe that every entropy of y for each value of x is no more than H(y). Moreover, the value got from formula (1) is not the entropy of y when x occurs, but the weighted average of conditional entropies for the various possible values of x.

## IV. Counterexamples for non-increasing of the entropy

In addition to the above analysis, the counterexamples below are listed to show that the conditional entropy may be increased.

Example 1: As the disciplines are strict in a certain school, the students can go to school before the designated time, with probability of 0.01 for them to be late for school. But Alice obtained the news from Bob that Rose who is in the school rarely complied with the discipline (including be late for school). The uncertainty of what Bob told Alice is eliminated when the news is known, but the uncertainty of Rose's being late for school or not is increased. For it could been speculated that Rose might often go to school in time, and the prior probability that she has not been late for school might be 0.99 as the probability of being late is only 0.01when we known nothing about Rose and the strict disciplines in this school stated above. When we got the message that Rose rarely complied with the discipline, it could be speculated that the posterior probability for Rose to be late for school would be increased. If the posterior probability for Rose to be late for school is less than 0.99, it can be found that the uncertainty of Rose's being late for school or not would be increased according to Shannon's theory. We can see that the posterior (conditional) entropy would be more than the prior entropy[7].

Example 2: The plaintext space is M = (0,1). According to the communication context, it is first known that the prior probability of plaintext being 0 is 0.9, while the prior probability of plaintext being 1 is 0.1. The ciphertext space is C = (0,1) and the key space is K = (0,1) and the keys are equally likely. The cryptoalgorithm is one time pad. Later the information is obtained that the ciphertext is 0. When the later information is considered, for the fixed ciphertext, there is a one-to-one correspondence between all the plaintexts and keys, so it can be concluded that the plaintexts are equally likely, that is, the probability of plaintext being 1 is 0.5. As the probability obtained above isn't consistent with the prior probability, the compromise is needed. The compromised posterior probability of the plaintext would be no more than the larger and no less than the smaller of the two corresponding probabilities of the two conditions. The uncertainty of the posterior probability is larger than uncertainty of the prior probability and the posterior entropy is increased, that is, the conditional entropy is increased.

## V. Conclusion

The paper analyzes the problems of definition and calculation of conditional entropy from various angles, and rectifies the calculation formula of conditional entropy. The increasing of conditional entropy is proved to be possible through illustrations. That indicates information is not fit to be defined as the thing which reduces uncertainty. I gave a new definition of information based on reliability of information after the analyses of the reliability and relativity of information[5,8]. The new definition shows much more concern to the reliability, which will make the information theory more consistent with the reality and adaptable to social needs. That will expand the application areas of information theory and promote the fusion of information, knowledge and intelligence [9].

- 

The Project Supported by Guangxi Science Foundation (0640171) and modern communication national key laboratory Foundation (No. 9140C1101050706)

biography：

**WANG Yong**(1977－),Hubei province tianmen city ,male,master of cryptography, research fields：cryptography,information security,Generalized Information Theory ,quantum information technology. GuiLin University Of Electronic Technology ,Guangxi, Guilin, 541004 E-mail: hellowy@126.com wang197733yong@sohu.com

Mobile (86)13978357217 tel：(86)7735603917 (home) fax：(86)7735601330(office)

School of Computer and Control, GuiLin University Of Electronic Technology, Guilin City, Guangxi Province, China, 541004